\documentclass[pre,floatfix,pdftex,twocolumn,ltxgrid]{revtex4}
\usepackage[pdftex]{graphicx} \usepackage{subfigure}
\usepackage{stdclsdv} \usepackage{array} \usepackage{amsmath}

\usepackage{balance} 
\usepackage[english]{babel}
\usepackage{graphicx} \usepackage{subfigure}
\usepackage{stdclsdv} \usepackage{array}
\usepackage{siunitx}
\usepackage{amsmath}
\usepackage{amsfonts}
\usepackage{amssymb}
\usepackage{comment}

\usepackage{bm}
\usepackage{graphicx} 
\usepackage{lastpage}
\usepackage[format=plain,justification=raggedright,singlelinecheck=false,font=small,labelfont=bf,labelsep=space]{caption} 

\newcommand{\dd}{\; \mathrm{d}}

\newcommand{\Av}[1]{\overline{ #1 }}
\newcommand{\Var}{\mathrm{Var}}

\newcommand{\Dl}{D}
\newcommand{\Dw}{\Delta}
\newcommand{\BStrain}{\gamma}

\newcommand{\bint}{\oint_{\partial\Omega}}
\begin{document}

\title{Connecting local active forces to macroscopic stress in elastic media}

\author{Pierre Ronceray}\email{pierre.ronceray@u-psud.fr}~\affiliation{Univ. Paris-Sud; CNRS; LPTMS; UMR 8626, Orsay 91405 France.} 
\author{Martin Lenz}\email{martin.lenz@u-psud.fr}~\affiliation{Univ. Paris-Sud; CNRS; LPTMS; UMR 8626, Orsay 91405 France.} 

\begin{abstract}
  In contrast with ordinary materials, living matter drives its own
  motion by generating active, out-of-equilibrium internal
  stresses. These stresses typically originate from localized active
  elements embedded in an elastic medium, such as molecular motors
  inside the cell or contractile cells in a tissue. While many
  large-scale phenomenological theories of such active media have been
  developed, a systematic understanding of the emergence of stress
  from the local force-generating elements is lacking. In this paper,
  we present a rigorous theoretical framework to study this
  relationship. We show that the medium's macroscopic active stress
  tensor is equal to the active elements' force dipole tensor per unit
  volume in both continuum and discrete linear homogeneous media of
  arbitrary geometries. This relationship is conserved on average in
  the presence of disorder, but can be violated in nonlinear elastic
  media. Such effects can lead to either a reinforcement or an
  attenuation of the active stresses, giving us a glimpse of the ways
  in which nature might harness microscopic forces to create active
  materials.
\end{abstract}

\maketitle


\section{Introduction} Forces in living systems are largely generated
at the nanometric protein level, and yet biological function often
requires them to be transmitted to much larger length scales. In the
actomyosin cytoskeleton for instance, local forces exerted by myosin
molecular motors on a disordered elastic scaffold of actin fibers
determine the mechanical properties of the cell and help drive
mitosis, cell migration and adhesion~\cite{alberts}. At a larger scale,
contractile cells exert forces on their surroundings to participate in
muscular contraction, clot stiffening~\cite{platelets} and wound
healing~\cite{Ehrlich:1988}. Due to their physiological relevance, such
systems have been extensively studied \emph{in vitro}, and direct,
dynamical imaging has recently progressed from macroscopic
observations to visualizations of the networks'
microstructure~\cite{bendix,Silva:2011} as well as individual
components ~\cite{Murrell:2012} during contraction.

The abundance of different macroscopic behaviors generated by
apparently similar microscopic components, which is particularly
spectacular in the cytoskeleton, has attracted significant theoretical
attention over the last decade. Two prominent theoretical strategies
have emerged.

On the one hand, so-called ``active gels'' models emphasize
macroscopic flows within the cytoskeleton, and do not formulate
detailed assumptions about the microscopic interactions between motors
and filaments ~\cite{Julicher:2007,joanny,marchetti-rmp}. Instead, they
rely on symmetry considerations to derive the most general equations
compatible with the problem considered, and successfully predict
intricate patterns of motion resembling experimentally observed
dynamical structures. While very general, these approaches involve a
large number of unprescribed parameters enclosing the relevant aspects
of the microscopic dynamics; in particular, the most important,
specifically active aspects of the cytoskeletal dynamics are typically
described by a phenomenological ``active stress tensor''.

On the other hand, length scales too small to be accurately captured
by an active gel formalism have typically been modeled using both
continuum~\cite{safran} and discrete~\cite{Head:2005aa} elastic models,
yielding insights into specific cellular processes such as mitotic
spindle organization~\cite{Loughlin:2010}, lamellipodium
growth~\cite{Atilgan:2005aa} or intracellular
propulsion~\cite{Zhu:2010}.  However, although the bulk elastic
properties of such models have been thoroughly
investigated~\cite{chase_review} on a general basis, force transmission
from the microscopic to the macroscopic level was only considered in
numerical simulations of specific
geometries~\cite{Head:2005aa,carlsson,dasanayake,nair,chase,Dasanayake:2013},
and a general theoretical framework to understand this process is
lacking.

\begin{figure*}[t]
  \includegraphics[width=\textwidth]{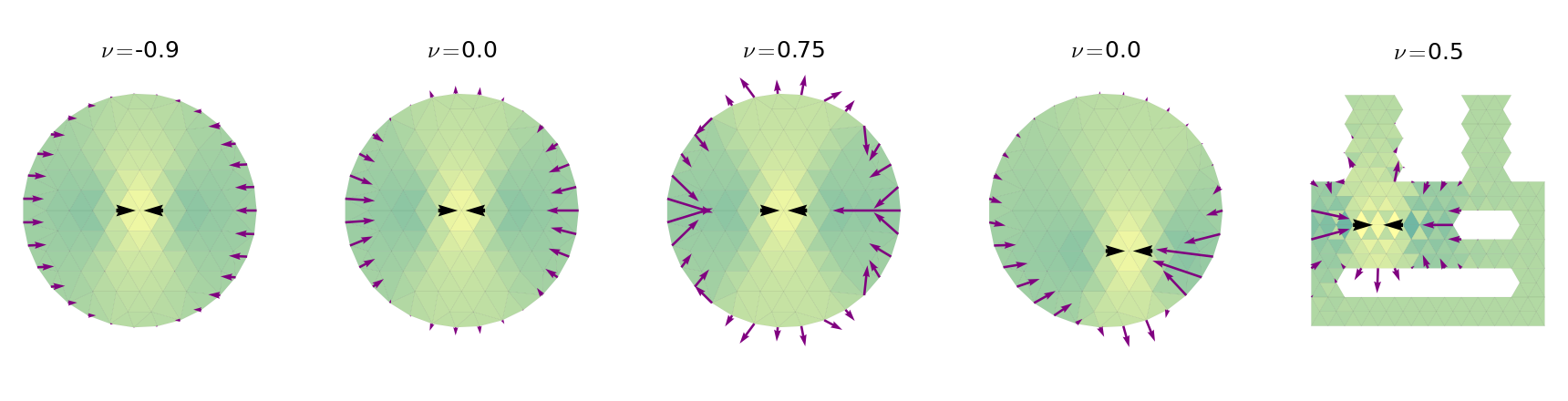}
  \caption{Boundary forces (\emph{purple arrows}) transmitted by a
two-dimensional homogeneous linear elastic medium under the influence
of a localized force dipole (\emph{black arrowheads}) computed using
finite elements. The boundary force distribution is strongly
influenced by both the medium's material properties ($\nu$ denotes the
Poisson ratio) and the geometry of the problem. Nevertheless, in all
cases the boundary dipole tensor is equal to the body forces' dipolar
moment.\label{fig:geometries}}
\end{figure*}

In this paper, we introduce such a formalism under the form of a
direct relation--termed ``dipole conservation''--between the
macroscopic active stress and the force dipole tensor, a local
quantity describing the individual force-exerting elements. Going
beyond previous special-case derivations, we show that this relation
applies in both continuum (Sec.~\ref{sec:continuum}) and discrete
(Sec.~\ref{sec:discrete}) homogeneous, linear elastic media
irrespective of their shape and of the spatial distribution of the
active forces. To understand the biologically relevant influence of
heterogeneities, we investigate the case of random spring networks in
Sec.~\ref{sec:disorder}, and show that although dipole conservation is
violated in individual realizations of the network it still holds in
an average sense provided the disorder is the same
everywhere. Finally, in Sec.~\ref{sec:NL_continuum} we study a toy
model nonlinear elastic medium and show that nonlinear elasticity can
skew force transmission towards either contraction or
extension. Sec.~\ref{sec:discussion} then discusses the applicability
of this result to existing models of force transmission in tissues and
the cytoskeleton.

\section{Dipole conservation in continuum elastic
media}\label{sec:continuum}

The transmission of localized active forces to the outer boundary of a
continuum elastic body is a geometrically complex problem, and the
distribution of transmitted forces strongly depend on the body's shape
and material properties (Fig.~\ref{fig:geometries}). Nevertheless,
here we show that strong nonlocal constraints exist between body and
boundary forces. In Sec.~\ref{sec:wall_stress} we introduce the
boundary dipole tensor, a quantity characterizing the boundary forces
that is directly related to the stress tensor. Using general
conditions of mechanical equilibrium, we relate this boundary dipole
to the spatial distribution of body forces in
Sec.~\ref{sec:MST}. Specializing our result to homogeneous linear
media, we then show in Sec.~\ref{sec:invariant} that the boundary
dipole is exactly equal to the dipolar moment of the body force
distribution, which we refer to as ``dipole conservation''.

\subsection{Boundary dipole tensor}
\label{sec:wall_stress} Let us consider a general $d$-dimensional
piece of elastic material at mechanical equilibrium, filling a domain
$\Omega$ of space with boundary $\partial \Omega$ and volume $V$. We
model the active elements embedded in the elastic body as a
distribution of body forces  $F_\mu(\mathbf{R})$. To quantify the macroscopic consequences of these active forces, we consider the response of the total system composed of the elastic medium and the embedded elements to an infinitesimal, affine
deformation characterized by a strain tensor $\BStrain_{\mu\nu}$. Under this transformation, a point belonging to the boundary $\partial \Omega$ of the elastic body with position $\mathbf{R}$ in the resting state
is displaced by a quantity $\delta \! R_\mu(\mathbf{R}) =
\BStrain_{\mu\nu}R_\nu$ (summation over repeated Greek indices is
implied)~\footnote{Throughout this article, the positions $\mathbf{R}$
  can be chosen to refer to either the undeformed or deformed state,
  provided that the correct form of the stress (nominal \emph{vs.}
  Cauchy) is used. The mean-stress theorem holds for arbitrary
  deformations in both cases.}. Denoting the elastic stress tensor by
$\sigma_{\lambda\mu}(\mathbf{R})$ and considering a surface element
$\dd s_{\lambda}$ lying on the boundary $\partial \Omega$, the force
exerted by the outside world on the surface element reads
$-\sigma_{\lambda\mu}(\mathbf{R}) \dd s_{\lambda}$. As the surface
element gets displaced by an infinitesimal $\delta \! R_\mu$, the work
performed by this force is $[-\sigma_{\lambda\mu}(\mathbf{R}) \dd
s_{\lambda}]\times\delta \! R_\mu$. The change in energy of the total
system is given
by the work performed over the whole boundary:
\begin{eqnarray} 
  \delta E &=& \bint [-\sigma_{\lambda\mu}(\mathbf{R})
  \dd s_{\lambda}]\times\delta \!  R_\mu \nonumber\\ &=& -
  \BStrain_{\mu\nu}\times \bint
  \sigma_{\lambda\mu}(\mathbf{R})R_\nu \dd
  s_{\lambda}.  \label{eq:affine}
\end{eqnarray} Noting that the integral in the right-hand side of
Eq.~(\ref{eq:affine}) is the dipolar moment of the boundary forces, we
refer to this quantity as the ``boundary dipole tensor'' and denote it
as
\begin{equation}
  \label{eq:Dwall_energetics} \Dw_{\mu\nu} = \bint
\sigma_{\lambda\mu}(\mathbf{R})R_\nu \dd s_{\lambda}.
\end{equation}

The meaning of this new quantity becomes clear if we note that
according to Eq.~(\ref{eq:affine}), $\Dw_{\mu\nu}$ is the derivative
of the energy of the total system with respect to the boundary strain
$\BStrain_{\mu\nu}$. This is reminiscent of the definition of the
stress tensor $\sigma_{\mu\nu}$ as the derivative of the energy
density $e$ with respect to the local strain tensor
$\gamma_{\mu\nu}(\mathbf{R})$. Considering a coarse-grained approximation of the total system with a uniform bulk deformation
$\BStrain_{\mu\nu}$ and uniform stress
$\tilde{\sigma}_{\mu\nu}$, we have $E=Ve$ with $e$
a uniform elastic energy density and the boundary dipole tensor is
directly related to the coarse-grained stress tensor:
\begin{equation}\label{eq:stress_tensor} \Dw_{\mu\nu} = \frac{\partial
(Ve)}{\partial \gamma_{\mu\nu}}=-V\tilde{\sigma}_{\mu\nu}.
\end{equation} 
Thus $-\Dw_{\mu\nu}/V$ is the medium's coarse-grained stress tensor
and $\Dw/(Vd) = \Dw_{\mu\mu}/(Vd)$ is the analog of a hydrostatic
pressure. In an active medium language, $\Dw<0$ thus characterizes a
contractile medium while $\Dw>0$ is associated with extensility.

Note that in a system with periodic boundary condition, the boundary dipole
tensor can be defined through the relation $\Dw_{\mu\nu} = -{\partial
E}/{\partial (\BStrain_{\mu\nu})}$, where the affine deformation
can be imposed through Lees-Edwards boundary conditions. Unless
explicitly stated, all the continuum and discrete results presented in
this manuscript can be rederived under periodic boundary conditions
with only minimal modifications to their proofs.

\subsection{Mean-stress theorem}
\label{sec:MST} 

As a first step towards establishing dipole conservation, here we
rederive a result known as the mean-stress
theorem~\cite{carlsson,gurtin}. We introduce the dipolar moment of the
active forces $F_\mu(\mathbf{R})$ as
\begin{equation}
  \label{eq:D_def} \Dl_{\mu\nu} = \int_\Omega F_\mu(\mathbf{R}) R_\nu
\dd V.
\end{equation} Note that $\Dl_{\mu\nu}$ is independent of the origin
of the coordinates if the body forces sum to zero as expected for
active elements embedded in an elastic medium, and that the total
force dipole exerted by several active elements is equal to the sum of
the individual force dipoles.

Inserting the force balance equation $\partial_\nu \sigma_{\mu\nu} =
-F_\mu$ into Eq.~(\ref{eq:D_def}) and integrating by part yields the
mean stress theorem
\begin{equation}\label{eq:IPP} \Dl_{\mu\nu} = \bint
\sigma_{\lambda\mu}(\mathbf{R}) R_\nu \dd s_{\lambda} + \int_{\Omega}
\sigma_{\mu\nu}(\mathbf{R}) \dd V.
\end{equation} Defining the integrated stress tensor $\Sigma_{\mu\nu} =
\int_{\Omega} \sigma_{\mu\nu} \dd V$ and using the definition
of the boundary dipole Eq.~(\ref{eq:Dwall_energetics}),
Eq.~(\ref{eq:IPP}) can be cast into a compact form:
\begin{equation}
    \label{eq:MST} \Dw_{\mu\nu}= \Dl_{\mu\nu}- \Sigma_{\mu\nu}.
  \end{equation} 
  This result holds irrespective of the medium's material properties,
  including homogeneity and linearity.

\subsection{Dipole conservation}
\label{sec:invariant} Despite its universality, in the general case
the result of Eq.~(\ref{eq:MST}) involves a complicated unknown object
$\Sigma_{\mu\nu}$ and is thus of limited practical use. Here we show
that this limitation is lifted when considering a linear homogeneous
elastic medium with fixed boundaries.

In a linear homogeneous elastic medium, stress is related to strain
through a position-independent stiffness tensor:
$\sigma_{\mu\nu}(\mathbf{R}) = C_{\mu\nu\alpha\beta}
\gamma_{\alpha\beta}(\mathbf{R})$. Integrating this relation over
space, we get
\begin{equation}
  \label{eq:stress_linhom} \Sigma_{\mu\nu} = C_{\mu\nu\alpha\beta} \Gamma_{\alpha\beta}
 \quad\textrm{with}\quad \Gamma_{\alpha\beta} = \int_\Omega \gamma_{\alpha\beta}(\mathbf{R}) \ \dd V,
\end{equation}
with $\Gamma_{\alpha\beta}$ the integrated strain. Reminding ourselves
that $\gamma_{\alpha\beta}(\mathbf{R})=[\partial_\alpha
u_\beta(\mathbf{R})+\partial_\beta u_\alpha(\mathbf{R})]/2$ with
$u_\alpha(\mathbf{R})$ the medium's displacement vector, integration
of Eq.~(\ref{eq:stress_linhom}) yields a boundary integral
\begin{equation}\label{eq:boundary_strain}  
  \Gamma_{\alpha\beta} =\bint\left[\frac{u_\beta(\mathbf{R})}{2}
    \dd s_\alpha+\frac{u_\alpha(\mathbf{R})}{2} \dd s_\beta\right].
\end{equation} 
Equation~(\ref{eq:MST}) thus provides a decomposition of the boundary
stress as a sum of a bulk term $\Dl_{\mu\nu}$ involving active forces
and a boundary term $\Sigma_{\mu\nu}= C_{\mu\nu\alpha\beta} \Gamma_{\alpha\beta}$ related to the system
deformation. Note that the latter depends on the system's elastic
properties through the stiffness tensor $C_{\alpha\beta\mu\nu}$, while
the former does not. Now introducing the assumption of fixed boundary
conditions, we find that the boundary displacements in the right-hand
side of Eq.~(\ref{eq:boundary_strain}) vanish, implying that the whole
integral vanishes. Using Eq.~(\ref{eq:stress_linhom}), we thus find
that $\Sigma_{\mu\nu}=0$, and thus Eq.~(\ref{eq:MST}) can be rewritten
as the dipole conservation relation:
\begin{equation}
  \label{eq:linhom} 
  \Dw_{\mu\nu} = \Dl_{\mu\nu}
\end{equation}
which relates bulk and boundary forces. To understand the meaning of
this equation, we decompose it into the equality of the traces,
symmetric traceless parts and antisymmetric parts of the two
tensors. The equality of the traces, $\Dw = \Dl_{\mu\mu}=\Dl$, is of
particular interest for biological systems as it relates the
``hydrostatic pressure'' $\Dw$ of the medium to the local force dipole
$\Dl$, a quantity routinely interpreted as the amount of contractility
of the active
elements~\cite{lenz1,dasanayake,carlsson,mackintosh,chase}. Next, the
symmetric traceless part of each of the two dipole tensors
[$(\Dw_{\mu\nu}+\Dw_{\nu\mu})/2$ and $(\Dl_{\mu\nu}+\Dl_{\nu\mu})/2$]
is analogous to a nematic order parameter characterizing the
anisotropy of the corresponding forces, and thus their equality means
that the anisotropy of the contractile forces is also conserved across
scales. Finally, the equality
$\Dw_{\mu\nu}-\Dw_{\nu\mu}=\Dl_{\mu\nu}-\Dl_{\nu\mu}$ of the
antisymmetric parts is equivalent to torque balance in the elastic
medium; since embedded active elements exert a vanishing total torque
on the elastic medium, it simply reduces to
$\Dw_{\mu\nu}-\Dw_{\nu\mu}=0$, and thus expresses torque balance on
the total system.

For systems without fixed boundaries, Eq.~(\ref{eq:linhom}) takes the more general form
\begin{equation}
\Dw_{\mu\nu} = \Dl_{\mu\nu} - C_{\mu\nu\alpha\beta} \Gamma_{\alpha\beta},
\end{equation}
meaning that the total coarse-grained stress $-\Dw_{\mu\nu}/V$ is the
sum of an active contribution and of the elastic stress $
C_{\mu\nu\alpha\beta} \Gamma_{\alpha\beta}$.

Note that Eq.~(\ref{eq:linhom}), as well as the other dipole
conservation relations presented in this paper assume a homogeneous
(or statistically homogeneous in Sec.~\ref{sec:disorder}) elastic
medium. Like these other results, it can however be generalized to
cases where a piece of elastic material is removed to make space for
the embedded active element by introducing a correction to the local
dipole accounting for the missing piece.

\section{Dipole conservation in discrete elastic media}
\label{sec:discrete} We now prove dipole conservation in discrete
media, with similar implications as in the continuum case of
Sec.~\ref{sec:continuum}. Although more technically involved, this new
derivation parallels the one of the previous section and its results
have a similar physical interpretation. We introduce the active force dipole tensor and the boundary
dipole tensor in Sec.~\ref{sec:discrete_Dw} and show that it satisfies
a discrete mean-stress theorem in Sec.~\ref{sec:DMST}. Dipole conservation
is then derived in Sec.~\ref{sec:discrete_invariants} under the
assumptions of linearity and local point reflection symmetry, a
variant of the homogeneity assumption used above.

\subsection{Active force and boundary dipole tensors}\label{sec:discrete_Dw} 
\begin{figure}[t]
  \includegraphics[width=\columnwidth]{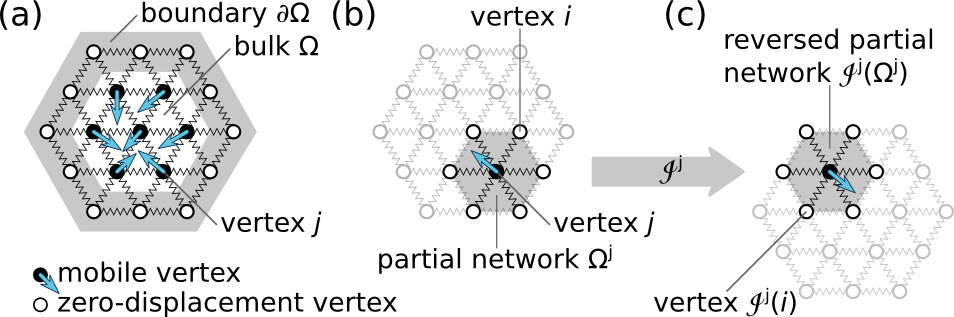}
  \caption{\label{fig:discrete}Parametrization and point reflection
invariance in a discrete elastic system (a)~Mobile bulk vertices (\emph{solid
circles}) comprised in the bulk $\Omega$ of the network are connected
to each other and to zero-displacement boundary vertices belonging to
the boundary $\partial\Omega$ (\emph{open circles}). \emph{Blue arrows} represent their displacements. (b)~The partial network $\Omega^j$ is obtained by setting all displacements to zero except that of vertex $j$. (c)~The partial network $\Omega^j$ is invariant under point reflection about vertex $j$ even though the total network $\Omega$ (\emph{in grey}) is not. The displacement of vertex $j$ is reversed under this transformation.}
\end{figure}

We consider
a $d$-dimensional system $\Omega$ comprised of interacting vertices
$i$ located at positions $\mathbf{R}^{(i)}$ in the reference
configuration, and at $\mathbf{R}^{(i)} + \mathbf{u}^{(i)}$ in the
deformed configuration characterized by the displacements
$\mathbf{u}^{(i)}$. The system's boundary $\partial \Omega$ consists
in a set of additional vertices whose displacements are set to zero
[see Fig.~\ref{fig:discrete}(a)]. The active force dipole tensor and the boundary dipole tensor are thus respectively defined as
\begin{subequations}
\begin{eqnarray}
\Dl_{\mu\nu} &=&
\sum_{i\in \Omega} F_\mu^{(i)}\ R_\nu^{(i)}, \label{eq:Dl_discrete}\\
\Dw_{\mu\nu} &=& \sum_{l\in \partial\Omega}
f_{\mu}^{(i)}\ R_{\nu}^{(i)} \label{eq:Dw_discrete}
\end{eqnarray}
\end{subequations} where $F_\mu^{(i)}$ is the body force
applied on the elastic network at vertex $i$ and $f_\mu^{(i)}$ is the force exerted by the
system on boundary vertex $i$.

\subsection{Mean-stress theorem}
\label{sec:DMST} As in the continuum case, the discrete mean-stress
theorem stems from force balance. Here we consider only forces between
pairs of vertices, as many-body interactions can always be decomposed
into sums of pair interactions. We assume these interactions to have
finite range. Denoting by $f_\mu^{(ij)}$ the force exerted by vertex
$i$ on vertex $j$, the force balance condition reads
\begin{subequations}
\begin{eqnarray} F_\mu^{(i)} &=& \sum_{j\sim i}
f_\mu^{(ij)} \label{eq:force_balance}\\ f_\mu^{(i)} &=& \sum_{j\sim i}
f_\mu^{(ij)} \label{eq:force_boundary}
\end{eqnarray}
\end{subequations} for bulk and boundary vertices, respectively. Here
$\sum_{j\sim i}$ denotes a sum over the vertices $j$ that interact
with $i$.

Inserting Eq.~(\ref{eq:force_balance}) into
Eq.~(\ref{eq:Dl_discrete}), we obtain a double sum over vertices of
the form $\sum_{i\in\Omega}\sum_{j\sim i}$. Reorganizing it into a sum
over pairs of neighboring vertices and splitting the resulting
expression into two sums, one over bulk pairs and the other over pairs
straddling the boundary, we use Newton's third law $f_\mu^{(ij)} =
-f_\mu^{(ji)}$ to find
\begin{equation}
  \label{eq:MST_discrete_1} \Dw_{\mu\nu} = \Dl_{\mu\nu} + \sum_{(ij)}
f_\mu^{(ij)}\left[R^{(j)}_\nu - R^{(i)}_\nu\right]
\end{equation}
where the sum runs over all pairs of interacting vertices, including
boundary vertices. Defining the stress associated with a pair of
interacting vertices as~\footnote{Several different conventions can be
  chosen to generalize the stress tensor to discrete
  systems~\cite{liu,zimmerman,egami}. Here we chose to localize the
  stress on the bonds of the network, yielding a mean-stress theorem
  with a concise expression.}
\begin{equation}
  \label{eq:stress-bond} \sigma_{\mu\nu}^{(ij)} = -
f_\mu^{(ij)}\left[R^{(j)}_\nu - R^{(i)}_\nu\right]
\end{equation} 
we obtain
\begin{equation}
  \label{eq:MST_discrete} \Dw_{\mu\nu} = \Dl_{\mu\nu} -
\Sigma_{\mu\nu}\quad\textrm{with}\quad \Sigma_{\mu\nu} =\sum_{(ij)}
\sigma_{\mu\nu}^{(ij)},
\end{equation} 
which constitutes the discrete mean-stress theorem.

\subsection{Dipole conservation}
\label{sec:discrete_invariants}
As in the continuum case, here we assume linear elasticity to demonstrate $\Sigma_{\mu\nu}=0$, implying dipole conservation. Linearity implies that $\Sigma_{\mu\nu}$ is a linear function of the set $\left\{u^{(i)}_\lambda\right\}_{i\in\Omega}$ of equilibrium vertex displacements, which are themselves unspecified functions of the active forces. Therefore, the integrated stress in the network can be decomposed into a sum over fictitious partial networks $\Omega_j$ where all displacements but that of vertex $j$ are set to zero [Fig.~\ref{fig:discrete}(a-b)]:
\begin{equation}
  \label{eq:decomposition_linear}
  \Sigma_{\mu\nu}\left(\left\{u_\lambda^{(i)}\right\}_{i\in\Omega}\right)= \sum_{j\in\Omega} \Sigma_{\mu\nu}^{\Omega_j}\left(u^{(j)}_\lambda\right),
\end{equation}
where $\Sigma_{\mu\nu}^{\Omega_j}$ is the integrated stress in partial network $\Omega_j$.

To demonstrate dipole conservation, we show that the partial integrated stress $\Sigma_{\mu\nu}^{\Omega_j}$ vanishes for all $j$ in networks invariant under local point reflection. Considering a specific partial network $\Omega_j$, we first note that reversing the vertex displacement also reverses the integrated stress by linearity:
\begin{equation}\label{eq:reversal}
\Sigma_{\mu\nu}^{\Omega_j}\left(-u^{(j)}_\lambda\right) = - \Sigma_{\mu\nu}^{\Omega_j}\left(u^{(j)}_\lambda\right).
\end{equation}

We next introduce the assumption that each partial network $\Omega_j$ is invariant under local point reflection about vertex $j$. The result of this transformation is illustrated in Fig.~\ref{fig:discrete}(c), and we denote the symmetric of vertex $i$ by $\mathcal{I}^{j}(i)$. Since stresses are proper tensors, the integrated stress is unchanged under this transformation:
\begin{equation}\label{eq:propertensor}
\Sigma_{\mu\nu}^{\mathcal{I}^j\left(\Omega_j\right)}\left(\mathcal{I}^{j}\left(u^{\mathcal{I}^{j}(j)}_\lambda\right)\right) = \Sigma_{\mu\nu}^{\Omega_j}\left(u^{(j)}_\lambda\right),
\end{equation}
meaning that the point-reversed image of a system under, \emph{e.g.}, overall compression is a system under the same amount of overall compression. Since vertex $j$ is its own image under this transformation, its displacement is reversed: 
\begin{equation}
\mathcal{I}^{j}\left(u^{\mathcal{I}^j(j)}_\lambda\right)=\mathcal{I}^{j}\left(u^{(j)}_\lambda\right)=-u^{(j)}_\lambda.
\end{equation}
Noting that local point reflection means that the partial network $\Omega_j$ is invariant under $\mathcal{I}^{j}$, \emph{i.e.}, $\mathcal{I}^j(\Omega^{j})=\Omega^{j}$, Eq.~(\ref{eq:propertensor}) becomes
\begin{equation}\label{eq:invariance}
\Sigma_{\mu\nu}^{\Omega_j}\left(-u^{(j)}_\lambda\right) = \Sigma_{\mu\nu}^{\Omega_j}\left(u^{(j)}_\lambda\right).
\end{equation}

Combining Eqs.~(\ref{eq:reversal}) and (\ref{eq:invariance}), we find that $\Sigma_{\mu\nu}^{\Omega_j}=0$ for any $j$, which we insert into Eqs.~(\ref{eq:MST_discrete}) and (\ref{eq:decomposition_linear}) to prove dipole conservation in the original, full network $\Omega$:
\begin{equation}
  \label{eq:discrete_linhom}
  \Dw_{\mu\nu} = \Dl_{\mu\nu}
\end{equation}
Although superficially different from the translational invariance
used in Sec.~\ref{sec:continuum}, our local point reflection symmetry
has a similar physical meaning. Indeed, it states that from any point of
observation, the elastic medium looks the same to two observers
looking in opposite directions. It is however more restrictive than
translational symmetry, as it does not apply to, \emph{e.g.}, the
honeycomb and diamond lattices---we do however believe that dipole
conservation could be established in these lattices by considering
discrete rotational symmetries. Local point reflection symmetry
is nevertheless fulfilled by most usual lattices, including the
triangular, square, simple-, face-centered- and body-centered-cubic
lattices, and thus the discrete dipole conservation relation
Eq.~(\ref{eq:discrete_linhom}) remains of wide practical
interest. Furthermore, in a regular lattice with periodic boundary
conditions translational invariance is sufficient to prove dipole
conservation (with a proof similar to that presented in
Sec.~\ref{sec:exact_periodic}).

\section{Average dipole conservation in random elastic media}
\label{sec:disorder} 
\begin{figure}[t]
  \includegraphics[width=\columnwidth]{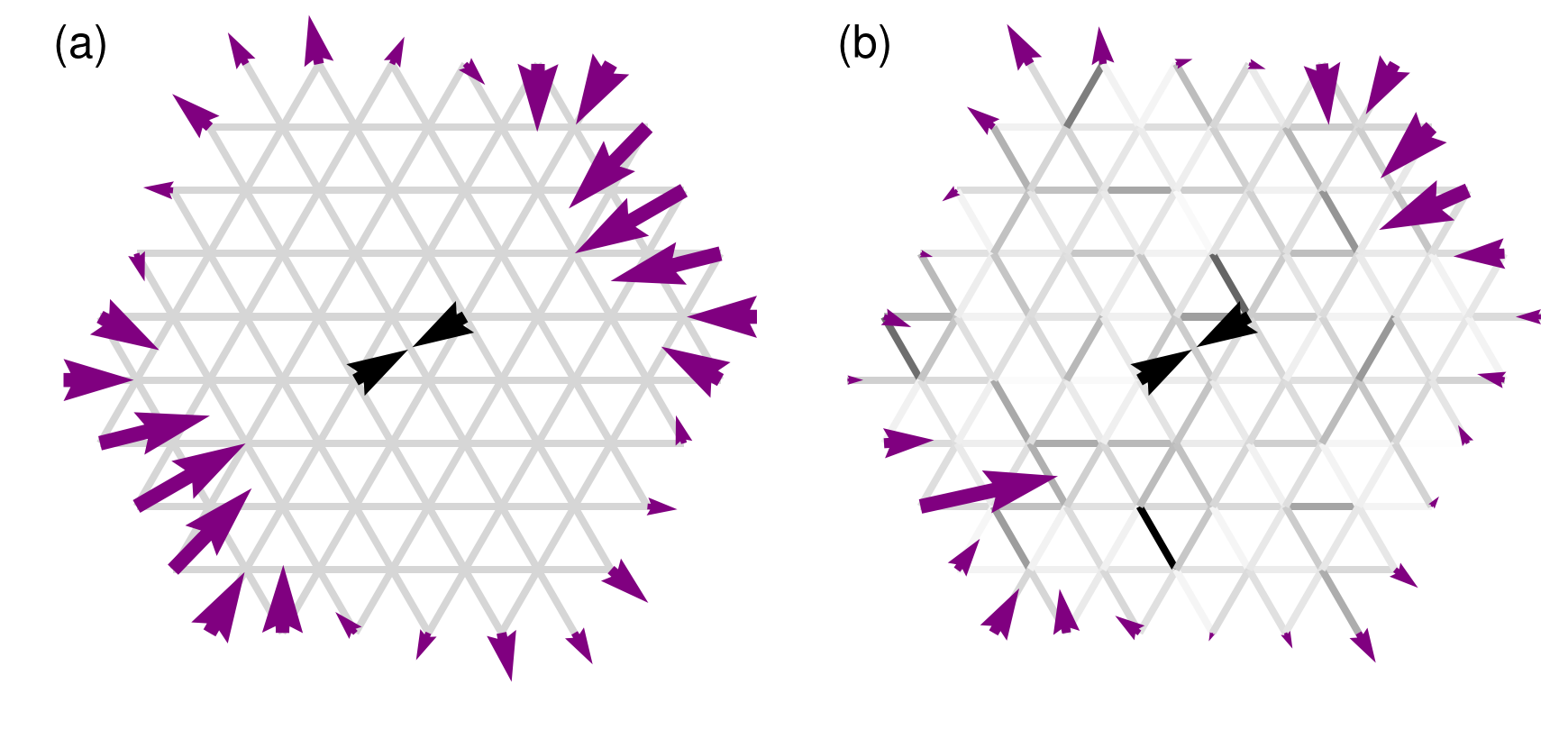}
  \caption{Force transmission in a linear spring network is strongly
affected by elastic inhomogeneities. Here the opacity of a bond is
proportional to its stiffness, and black arrowheads (purple arrows)
represent body (boundary) forces. (a)~In a homogeneous network, dipole
conservation ${\Dw} = \Dl$ is satisfied to the numerical
precision. (b)~In a random spring network, dipole conservation is
typically violated; in this specific example, ${\Dw}/\Dl \approx 0.60
$. Here the spring constants are drawn from a lognormal law with
standard deviation $\delta\alpha = 0.8$.\label{fig:disorder_network}}
\end{figure}

In this and the next section, we investigate how
relaxing the assumptions of homogeneity and linearity respectively
affect dipole conservation. As shown in
Fig.~\ref{fig:disorder_network}, inhomogeneous elastic properties
significantly affect dipole transmission in a spring
network. Nevertheless, we show in Sec.~\ref{sec:exact_periodic} that
in a random spring network with periodic boundary conditions dipole
conservation is preserved in an average
sense. Sec.~\ref{sec:disorder_numerics} then shows numerically that
fixed boundary conditions spoil this result, although deviations from
it are small and go to zero for large-size systems. Finally, in
Sec.~\ref{sec:EMT-results} we use an effective medium (\emph{i.e.},
mean-field) approach to quantify the sample-to-sample variations in
the amount of transmitted force dipole, and find that it is
proportional to the amplitude of the local spring disorder.

\subsection{Average dipole conservation in periodic geometry}
\label{sec:exact_periodic} Consider the linear response of a regular
lattice of independent, identically distributed random springs with
periodic boundary conditions subjected to a distribution of body
forces $F_\mu^{(i)}$ of zero sum (\emph{i.e.}, $\sum_{i}
F^{(i)}_\mu=0$ as expected for active elements embedded in an elastic
medium). Let $G^{(i)}_{\mu\nu\alpha}$ be the sample-dependent linear
response function relating the body force at site $i$ to the
integrated stress:
\begin{equation}
  \label{eq:linear_Sigma_force} \Sigma_{\mu\nu} = \sum_i
G^{(i)}_{\mu\nu\alpha} F^{(i)}_\alpha.
\end{equation} Denoting averages over lattice disorder by a bar, this
equation implies
\begin{equation}\label{eq:average_periodic_disordered_sigma}
\Av{\Sigma}_{\mu\nu} = \sum_{i} \Av{G^{(i)}_{\mu\nu\alpha}}
F^{(i)}_\alpha = \Av{G_{\mu\nu\alpha}} \sum_{i} F^{(i)}_\alpha,
\end{equation} where the second equality stems from the statistical
equivalence of all sites in the network, implying that the average
response function $\Av{G^{(i)}_{\mu\nu\alpha}}$ is independent of
$i$. Finally, inserting our assumption of vanishing sum of the body
forces into Eq.~(\ref{eq:average_periodic_disordered_sigma}) yields
$\Av{\Sigma}_{\mu\nu}=0$, and thus through Eq.~(\ref{eq:MST_discrete})
the force dipole is conserved on average:
\begin{equation}\label{average_dipole_conservation}
\Av{\Dw}_{\mu\nu}={\Dl}_{\mu\nu}.
\end{equation}
This result is valid in any system where all vertices are equivalent,
and thus also holds in infinite lattices.

\subsection{Violations of average dipole conservation in the presence
  of fixed boundaries}
\label{sec:disorder_numerics} 

To investigate the influence of finite domain size on the average
dipole conservation Eq.~(\ref{average_dipole_conservation}), we
numerically study the linear response to a force dipole of a finite
hexagonal system with fixed boundary conditions, as pictured in
Fig.~\ref{fig:disorder_network}(b). The network is a two-dimensional
triangular lattice of independent identically distributed random
hookean springs of unit rest length. The spring constant of the bond
joining two neighboring sites $i$ and $j$ is denoted $\alpha^{(ij)}$
and drawn from a distribution $\dd P(\alpha)$ with average
$\Av\alpha=1$ and variance $\Var(\alpha)=\delta\alpha^2$.

\begin{figure}[t]
  \includegraphics[width=\columnwidth]{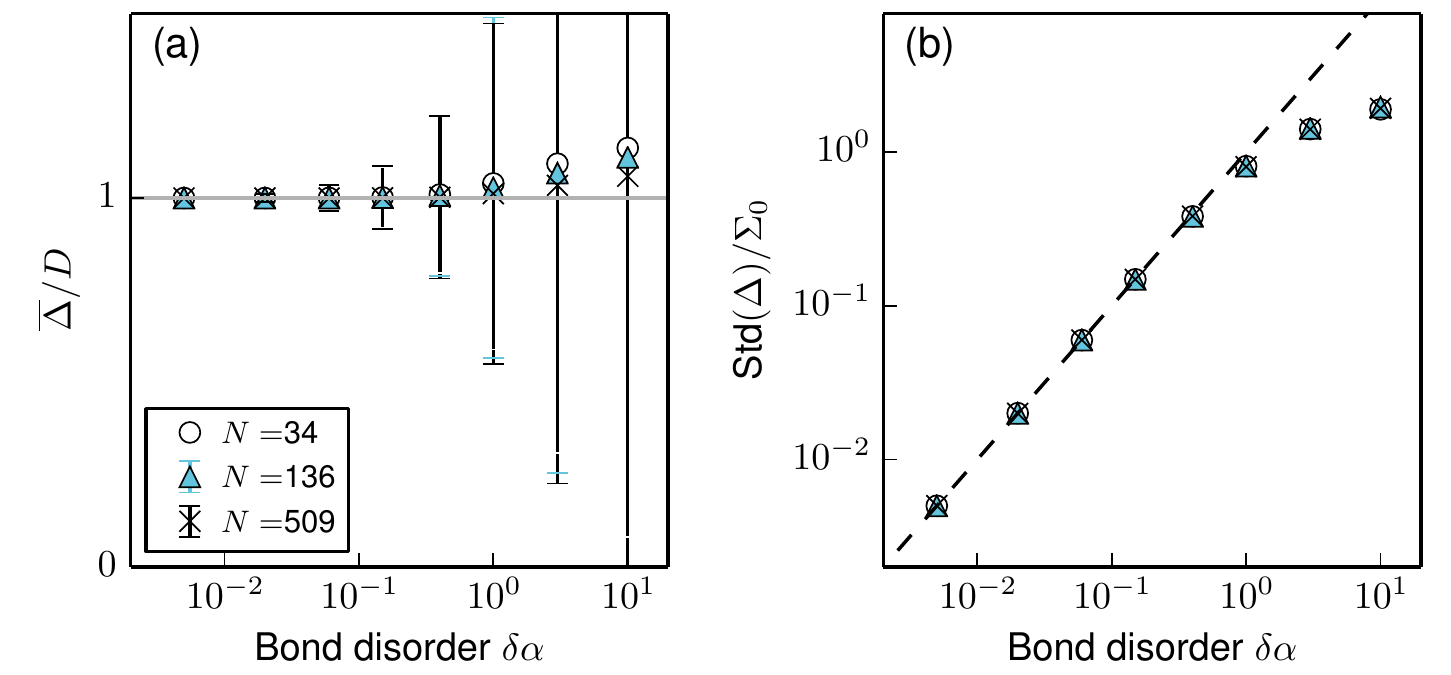}
  \caption{\label{fig:disorder_stats}Deviations from average dipole
    conservation and sample-to-sample fluctuations in random spring
    networks. (a)~The average dipole conservation condition
    $\Av\Dw/\Dl=1$ (\emph{grey line}) is well respected for systems
    with large enough number of mobile vertices $N$.  \emph{Bars}
    represent the standard deviation of this ratio, thus indicating
    the magnitude of sample-to-sample fluctuations. Each point in this
    figure represents data averaged over ${\cal O}(10^5)$ samples,
    ensuring that the plotted deviations in the average $\Av\Dw$ are
    statistically significant. (b)~Standard deviation of the boundary
    force dipole, $\Dw$ [proportional to the length of the bars in
    panel (a)] normalized by the second moment of the effective medium
    stress $\Sigma_0$ (see Appendix~\ref{sec:EMT}) as a function of
    disorder. We find good agreement with the small-disorder effective
    medium theory prediction Eq.~(\ref{eq:delta_sigma}) (\emph{dashed
      line}) up to $\delta\alpha\simeq 1$.}
\end{figure}

Assuming a lognormal form for $\dd P(\alpha)$, we minimize the elastic
energy of systems of different sizes using a conjugate gradient
algorithm. Our procedure uses displacements of order $10^{-100}$ times the lattice constant, thus guaranteeing that nonlinear effects are not
present in our results. Fig.~\ref{fig:disorder_stats}(a) shows that
average dipole conservation is violated for small systems, but that
these violations vanish for larger system sizes.

\subsection{Influence of network disorder on the reliability of force
transmission}
\label{sec:EMT-results} While in large enough systems the boundary
dipole becomes equal to the local force dipole \emph{on average},
Fig.~\ref{fig:disorder_stats}(a) shows that significant fluctuations
\emph{around} this average subsist even in infinite
systems. Physically, this stems from the fact that the configuration
of the immediate surroundings of the force-exerting active element can
strongly amplify or attenuate the local force dipole. These near-field
distortions are then faithfully propagated to long distances by the
more distant regions of the network, which tend to transmit forces in
a more dipole-conserving way. Therefore, due to their local origin
these distortions are not cured by increasing the system size. The
resulting boundary dipole fluctuations have a clear practical
significance, as they represent an intrinsic limitation on the
reliability of force transmission through disordered elastic networks
and thus represent a challenge for biological systems.

To better understand the magnitude of this effect in relation to the
amount of network disorder, we compute a mean-field-type approximation
of the boundary dipole fluctuations through an effective medium
theory~\cite{Feng}. As detailed in Appendix~\ref{sec:EMT}, effective
medium theories assimilate the effect of bond disorder in a fully
random network to that of a single random bond with spring constant
$\alpha$ immersed in an effective regular network. The spring constant
$\alpha_m$ associated with this effective network is chosen so that
the average of the displacement $v$ of the random bond in the regular
network is equal to the elongation $v_m$ of the non-random bonds,
\emph{i.e.}, $\Av{v}=v_m$. This formalism allows us to calculate an approximation of the
tension of each random bond, allowing us to compute the integrated
stress $\Sigma$. We find that the tension of the random bond is equal
to the bond tension in a fully regular medium plus a quantity
proportional to $v-v_m$. Since the integrated stress in the ordered
medium vanishes [Eq.~(\ref{eq:discrete_linhom})], our approximate
system has $\Sigma\propto v-v_m$. Now averaging this relation and
using $\Av{v}-v_m=0$, we find
\begin{equation}
  \label{eq:EMT-conservation} \Av{\Sigma} = \Dl-\Av{\Dw} = 0,
\end{equation} 
\emph{i.e.}, the effective medium theory predicts average dipole
conservation irrespective of boundary conditions.  Going beyond this
vanishing average stress, we further compute the variance
$\Av{\Sigma^2}$ of the integrated stress, which is proportional to
$\Av{(v-v_m)^2}$. For small disorder, the typical mismatch $v-v_m$
between the random bond and its deterministic neighbors is moreover
proportional to the mismatch $\alpha-\alpha_m$ of their spring
constants, and thus to the amplitude $\delta\alpha$ of the
disorder. This finally yields
\begin{equation}\label{eq:delta_sigma}
\textrm{Std}(\Sigma)  =\textrm{Std}(\Dw)
= \Sigma_0\delta\alpha,
\end{equation} 
where the geometry-dependent prefactor $\Sigma_0$ in the right-hand side is given in
Appendix~\ref{sec:EMT}. Comparing this effective medium prediction
with the numerical data of Sec.~\ref{sec:disorder_numerics}, we find
an excellent agreement up to a bond disorder $\delta\alpha\simeq 1$, following which our small-disorder expansion breaks down [Fig.~\ref{fig:disorder_stats}(b)].


This proportionality of dipole fluctuations $\delta\Sigma$ to the
network disorder $\delta\alpha$ suggests that reliable dipole
transmission is only possible in well-ordered media. However, due to
the linearity of the elastic medium, the fluctuations stemming from
many small contractile elements scattered through space average out to
zero. This scattered geometry is reminiscent of the structure of
force-generating cytoskeletal networks.

\section{Breakdown of dipole conservation in nonlinear elastic media}
\label{sec:NL_continuum} Unlike the elastic disorder discussed
above, nonlinear elastic behavior introduces systematic violations of
force dipole conservation, as illustrated here on a simple example. We
consider a spherical, three-dimensional cavity of radius $R_1$ filled
with a continuum homogeneous elastic medium with elastic energy
density
\begin{equation}
  \label{eq:strain-energy-NL} e = \frac{\lambda}{2}
\left(\mathrm{Tr}\gamma\right)^2 + \mu
\mathrm{Tr}\left(\gamma^2\right) + \frac{\beta}{3}
\left(\mathrm{Tr}\gamma\right)^3,
\end{equation} where $\gamma$ is the strain tensor, $\lambda$ and
$\mu$ are Lam\'e parameters that characterize the linear response of
the material, and $\beta$ is a nonlinear compressibility, with
$\beta>0$ describing softening upon compression. We impose a radial
displacement $u_0$ at radius $R_0$, resulting in a radial displacement
\begin{equation}
  \label{eq:NL-general_sol} u_R(R) = AR + \frac{B}{R^2}
\end{equation} with
\begin{subequations}
\begin{eqnarray} A = &
  \begin{cases} u_0/R_0 & R<R_0 \\ - u_0 R_0^2 / (R_1^3 - R_0^3) & R_0
< R < R_1
  \end{cases} \\ B = &
  \begin{cases} 0 & R<R_0 \\ u_0 R_0 R_1^3 / (R_1^{3} - R_0^{3}) & R_0
< R < R_1
  \end{cases}.
\end{eqnarray}
\end{subequations} Although Eq.~(\ref{eq:NL-general_sol}) matches the
linearized solution of the elastic problem, it is actually valid to
arbitrary nonlinear order for the specific form of the strain energy
of Eq.~(\ref{eq:strain-energy-NL})~\cite{horgan1984,horgan1988}.

Restricting ourselves to small displacements, we can use the usual
Cauchy strain tensor $\gamma_{\mu\nu} = \frac{1}{2}(\partial_\mu u_\nu
+ \partial_\nu u_\mu)$ and derive the resulting radial stress
\begin{equation}
  \label{eq:stress-NL} \sigma_{RR}(R) = 3 \lambda A + 2\mu \left(A -
\frac{2B}{R^3} \right) + 9 \beta A^2,
\end{equation} which we use to compute the local and boundary force
dipoles
\begin{eqnarray} \Dl &=& 4\pi R_0^3 \left[ \sigma_{RR}(R_0^+) -
\sigma_{RR}(R_0^-)\right] \nonumber\\ &=& u_{0} \left( \lambda + 2
\mu\right) \frac{12 \pi R_{1}^{3} R_{0}^{2} }{ R_{ 1}^{3} - R_{0}^{3}}
+ \beta u_{0}^{2} \frac{36 \pi R_{1}^{3} R_{0} \left( R_{1}^{3} - 2
R_{0}^{3}\right)} {\left(R_{1}^{3} - R_{0}^{3}\right)^2}\nonumber\\
{\Dw} & = & 4\pi R_1^3 \sigma_{RR}(R_1)\nonumber\\ &=& u_{0} \left(
\lambda + 2 \mu\right) \frac{12 \pi R_{1}^{3} R_{0}^{2} }{ R_{ 1}^{3}
- R_{0}^{3}} - \beta u_{0}^{2} \frac{36\pi R_{1}^{3} R_{0}^{4}
}{\left(R_{1}^{3} - R_{0}^{3}\right)^2}\nonumber\\ &\underset{R_0\ll
R_1}{\sim}&\Dl - 36\pi \beta u_{0}^{2} R_{0}.
\end{eqnarray} Thus the nonlinear elasticity of the material
renormalizes the local force dipole by a quantity $- 36\pi \beta
u_{0}^{2} R_{0}$ which becomes negligible in the linear limit $\beta
u_0 \ll \lambda,\mu$. This violation of dipole conservation favors
contraction ($\Dw<0$) for a material that softens under compression
($\beta>0$) as further discussed below.

\section{Discussion}
\label{sec:discussion} Stress-generating, active materials are
essential constituents of the cell, and their biological design is
strongly constrained by the physical laws governing force transmission
in elastic media. As shown here, these laws take a simple,
geometry-independent form in homogeneous linear elastic media, whereby
the force dipole is an invariant of linear elasticity. More
specifically, the macroscopic force dipole tensor exerted by the
medium on its boundaries is equal to the sum of the microscopic force
dipoles exerted on it by embedded active elements. This dipole
conservation relation is valid both for continuum media and for
discrete media with unspecified finite range interactions, making it
relevant for popular biological fiber network models with stretching
and bending energies~\cite{chase_review}. It also holds true in
anisotropic media. Due to its generality, dipole conservation is a
powerful tool to relate widely used macroscopic descriptions of the
cytoskeleton, sometimes termed active gels theories, to the underlying
microscopic phenomena. For instance, in a homogeneous linear elastic
medium with a density $\rho$ of embedded elements each exerting a
force dipole $d_{\mu\nu}$, the active stress $\sigma_{\mu\nu}$---the
central object of active gel theories---is simply given by
$\sigma_{\mu\nu}=-\rho d_{\mu\nu}$ [see Eqs.~(\ref{eq:stress_tensor})
and (\ref{eq:linhom})].

Considering more biologically relevant, disordered elastic media, we
show that in a discrete linear system where the disorder probability
distribution is position-independent, dipole conservation is satisfied
on average. This result again applies to fiber network models, but can
be violated in small systems where the influence of the boundary
conditions is not negligible. Dipole conservation is moreover not
generally respected in every statistical realization of the system,
and fluctuations are proportional to the amplitude of the disorder;
they are however self-averaging, leading to reliable, deterministic
stress generation in large enough systems.

Unlike disorder, nonlinearities have a systematic effect on force
transmission. Indeed, we show that a material that softens under
compression tends to favor contraction, reminiscent of the enhanced
contractility observed in bundles and networks of filaments prone to
buckling under compressive
stresses~\cite{Lenz:2012a,Lenz:2012,Notbohm:2014}. A similar effect has
been been predicted in shear stiffening materials~\cite{safran}. It is
moreover worth keeping in mind that nonlinear behavior in elastic
materials is not limited to constitutive nonlinearities in the
material properties, as nonlinear elasticity can also stem from
geometrical effects~\cite{Landau:1986aa}.  Importantly, such
geometrical nonlinearities are more prevalent in disordered than
homogeneous networks~\cite{Sheinman:2012}, implying that disorder might
significantly affect contractility by lowering the threshold to
nonlinear behavior. As a result, a reliable understanding of
contraction in active biological materials requires a good
characterization of the nonlinear property of the underlying elastic
matrix. Given impressive recent experimental advances in this area, we
believe that model-independent, rigorous theoretical studies such as
this one will be valuable in analyzing new data and thus understanding
the relation between molecular motors and cell-wide force generation.



\footnotesize{

\subsection*{Acknowledgements}

We thank Anders Carlsson, Chase Broedersz and Samuel Safran for fruitful discussions
and useful comments. Our group belongs to the CNRS consortium
CellTiss. This work was supported by grants from Universit\'e
Paris-Sud and CNRS, the University of Chicago FACCTS program, Marie
Curie Integration Grant PCIG12-GA-2012-334053 and ``Investissements
d'Avenir'' LabEx PALM (ANR-10-LABX-0039-PALM). PR is supported by
``Initiative Doctorale Interdisciplinaire 2013'' from IDEX
Paris-Saclay. Figures realized with
Matplotlib~\cite{matplotlib}.

\bibliographystyle{rsc} 

\appendix
\section{Effective medium theory for disordered spring networks}
\label{sec:EMT} Here we derive the results of
Sec.~\ref{sec:EMT-results} by developing an effective medium theory,
following Ref.~\cite{Feng}. In this approach the disordered network
described in Sec.~\ref{sec:disorder} is approximated by an effective
homogeneous network where every bond has a spring constant
$\alpha_m$. When subjected to the same body forces and boundary
conditions as the original network, the effective network deforms so
that the bond joining adjacent vertices $i$ and $j$ has elongation
$v^{(ij)}_m$ with respect to its rest length. To determine the value
of $\alpha_m$, we introduce a third system obtained by replacing bond
$(ij)$ by a random spring with constant $\alpha$ drawn with
probability law $\dd P(\alpha)$. This induces a change in the
deformation field, and the elongation of the considered bond in the
single-random-bond system is denoted $v^{(ij)} = v_m^{(ij)} + \delta
v^{(ij)} $. Mechanical equilibrium then imposes
\begin{equation}
  \label{eq:du} \delta v^{(ij)} = v_m^{(ij)} \frac{\alpha_m -
\alpha}{q \alpha_m + \alpha}
\end{equation} where $q = z/2d -1$ with $z$ the network connectivity
and $d$ the spatial dimension. The effective spring constant
$\alpha_m$ is fixed by imposing
\begin{equation}
  \label{eq:self_consistency} \Av{\delta v^{(ij)}} = v_m^{(ij)} \int
\frac{\alpha_m - \alpha}{q \alpha_m + \alpha} \dd P(\alpha) = 0,
\end{equation} where the average is taken over the distribution of
$\alpha$.

To compute the integrated stress $\Sigma$, we note that displacements
in our single random bond system are the same as in a homogeneous
lattice of $\alpha_m$ springs with an active force dipole of amplitude
$(\alpha-\alpha_m)v^{(ij)}$ applied along bond $(ij)$. We further note
that the integrated stress in this homogeneous, linear system vanishes
according to Eq.~(\ref{eq:discrete_linhom}). Since stresses in this
system are identical to those in our single-random-bond system except
at bond $(ij)$, the integrated stress in the latter is equal to the
integrated stress in the former (\emph{i.e.}, zero) plus the
contribution of bond $(ij)$:
\begin{equation}
  \label{eq:Sigma_EMT} \Sigma = 0+(\alpha - \alpha_m) v^{(ij)}=
\frac{z\alpha_m}{2d} \delta v^{(ij)}
\end{equation} where Eq.~(\ref{eq:du}) was used. Inserting
Eq.~(\ref{eq:self_consistency}) into Eq.~(\ref{eq:Sigma_EMT}), we
obtain $\Av{\Sigma}=0$, \emph{i.e.}, the average dipole conservation
equation Eq.~(\ref{eq:EMT-conservation}). 

Denoting $\sigma_m^{(ij)}=\alpha_m v_m^{(ij)}$ and
$\delta\sigma^{(ij)}=\sigma^{(ij)}_m+\delta\sigma^{(ij)}$, we plug
Eq.~(\ref{eq:du}) into Eq.~(\ref{eq:Sigma_EMT}) and compute
$\Av{\left[\delta\sigma^{(ij)}\right]^2} =
C\left[\sigma_m^{(ij)}\right]^2$, where
\begin{equation} C=\int \left[\frac{\alpha_m - \alpha}{(1-2d/z)
\alpha_m + 2d \alpha/z}\right]^2 \dd P(\alpha).
\end{equation} In the spirit of the effective medium theory, we
approximate the fully random lattice as a superposition of single
random bond lattices and sum the bond stresses $\sigma^{(ij)}$ as
independent identically distributed variables:
\begin{equation}
  \label{eq:varSigma} \Av{\Sigma^2} = \sum_{(ij)}
\Av{\left[\delta\sigma^{(ij)}\right]^2}= C \Sigma_0^2,
\end{equation} where $\Sigma_0^2=\sum_{(ij)}
\left[\sigma_m^{(ij)}\right]^2$ can be computed from the stress field
in the homogeneous system with appropriate boundary conditions and
active body forces. This procedure is used to obtain the normalization
factor of Fig.~\ref{fig:disorder_stats}(b). Note that $C$ takes a
simple form in the weak disorder limit
$\textrm{Var}(\alpha)=\delta\alpha^2\ll \Av{\alpha}^2$. Indeed,
setting $\Av{\alpha}=1$ Eq.~(\ref{eq:self_consistency}) yields
\begin{equation}
  \label{eq:small_disorder_alpha_m} \alpha_m = 1 - \frac{ 2d}{z}
(\delta\alpha)^2 + {\cal O}\left[(\delta\alpha)^3\right],
\end{equation} and the numerical factor becomes $C = \delta\alpha^2 +
{\cal O}\left[(\delta\alpha)^3\right]$, yielding
Eq.~(\ref{eq:delta_sigma}).

\end{document}